\newif\ifsubmode
\submodefalse
 
\ifsubmode
  \documentclass[12pt,preprint]{aastex}  
  \received{}
  \revised{}
  \accepted{}
 
\else
  \documentclass{emulateapj}  
  \slugcomment{Accepted to \apjl}
\fi

\newcommand{\kms}{km~s$^{-1}$}
\newcommand{\lsim}{\hbox{ \rlap{\raise 0.425ex\hbox{$<$}}\lower 0.65ex\hbox{$\sim$} }}
\newcommand{\gsim}{\hbox{ \rlap{\raise 0.425ex\hbox{$>$}}\lower 0.65ex\hbox{$\sim$} }}

\shorttitle{High-Redshift Supercluster}
\shortauthors{Gal \& Lubin}

\begin{document}
\vskip -1.0cm
\title{Spectroscopic Confirmation of the Cl 1604 Supercluster at $z
\sim 0.9$}

\author{R.R. Gal\altaffilmark{1} \& L.M. Lubin\altaffilmark{2}}
\affil{Department of Physics, University of California -- Davis, One
Shields Avenue, Davis, CA 95616}
\altaffiltext{1}{gal@physics.ucdavis.edu}
\altaffiltext{2}{lmlubin@ucdavis.edu}

\begin{abstract}
We present spectroscopic confirmation of the Cl 1604 supercluster at
$z \sim 0.9$. Originally detected as two individual clusters, Cl
1604+4304 at $z = 0.90$ and Cl 1604+4321 at $z = 0.92$,
which are closely separated in both redshift and sky position, subsequent
imaging revealed a complex of red galaxies bridging the two clusters,
suggesting that the region contained a large scale structure. We have
carried out extensive multi-object spectroscopy, which, combined with
previous measurements, provides $\sim600$ redshifts in this area,
including 230 confirmed supercluster members. We detect two
additional clusters that are part of this structure, Cl 1604+4314 at
$z = 0.87$ and Cl 1604+4316 at $z = 0.94$.  All four have properties
typical of local clusters, with line-of-sight velocity dispersions
between 489 and 962~{\kms}. The structure is significantly extended in
redshift space, which, if interpreted as a true elongation in real
space, implies a depth of $93~h_{70}^{-1}~{\rm Mpc}$. We examine the
spatial and redshift distribution of the supercluster members.
\end{abstract}


\section{Introduction}

The Cl 1604 supercluster was initially detected as two separate
clusters, Cl 1604+4304 at $z=0.90$ and Cl 1604+4321 at $z=0.92$, in
the plate-based survey of \citet{gho86}. Deeper imaging and
multi-object spectroscopy taken with the Low Resolution Imaging
Spectrograph \citep[LRIS;][]{oke95} at the Keck telescopes by
\citet[hereafter O98]{oke98} yielded redshifts and preliminary
velocity dispersions for each cluster \citep[][hereafter
P98,P01]{pos98,pos01}. Motivated by the close separation of Cl
1604+4304 and Cl 1604+4321 in both radial velocity (4300~{\kms}) and
position on the sky ($17'$), \citet{lub00} performed deep multi-band
imaging with the Palomar 5-m telescope covering the area between the
two clusters. The imaging revealed an overdensity of red galaxies
whose colors were consistent with early-type galaxies at $z \sim 0.9$,
suggesting that the clusters were part of a high-redshift
supercluster.

To verify this conclusion, map the supercluster structure, and obtain
a large sample of cluster galaxies with measured spectroscopic
properties, we are conducting an extensive spectroscopic survey
spanning the region between Cl 1604+4304 and Cl 1604+4321. In this
Letter, we present spectroscopic confirmation of this large scale
structure and a discussion of its properties based on
the currently known supercluster members. We assume a
$\Lambda$CDM cosmology with $\Omega_m=0.3, \Lambda=0.7,$ and ${\rm
H_0}=70~h_{70}~{\rm km~s^{-1}~Mpc^{-1}}$.

\section{The Spectroscopic Survey}

Here we briefly describe the spectroscopic survey of the Cl 1604
supercluster. A full discussion of the data acquisition and analysis
procedures will be presented in the catalog paper (Gal, Lubin \& Oke
2004, in preparation). The survey was conducted using LRIS and the
Deep Imaging Multi-object Spectrograph \citep[DEIMOS;][]{fab03} on the
Keck 10-m telescopes. Target selection was based on multi-band imaging
\citep{lub00} taken at the Palomar 5-m telescope.  A contiguous area
of $10.4' \times 18.2'$ (4.8 Mpc $\times 8.5$ Mpc) between the original two clusters was imaged
in $B$, $R$, and Gunn $i$ filters. The galaxy catalog reaches a depth
of $B\simeq26$, $R\simeq25$ and $i\simeq24$ for a $5\sigma$
detection. The resulting color-magnitude diagrams reveal a
well-delineated red sequence of galaxies whose colors ($1.2 \le R-i \le 1.7$) are consistent
with spectroscopically-confirmed, early-type galaxies in the two
original clusters \citep[see Figures 2 \& 4 of][]{lub00}. In
Figure~\ref{pal}, we indicate on the composite $i$ band image all of
these galaxies by overlaying contours of red  galaxy density. 
These contours clearly delineate the large system of galaxies encompassing
the original clusters. We also observe two strong concentrations of
red galaxies directly between Cl 1604+4304 and Cl 1604+4321,
suggesting the presence of two additional clusters. Objects within the
red sequence were preferentially (but not exclusively) targeted for
slit assignment in our current spectroscopic survey.  Targets were
observed with a combination of LRIS and DEIMOS slitmasks.

\ifsubmode
\else
\begin{figure}
\epsscale{.75}
\plotone{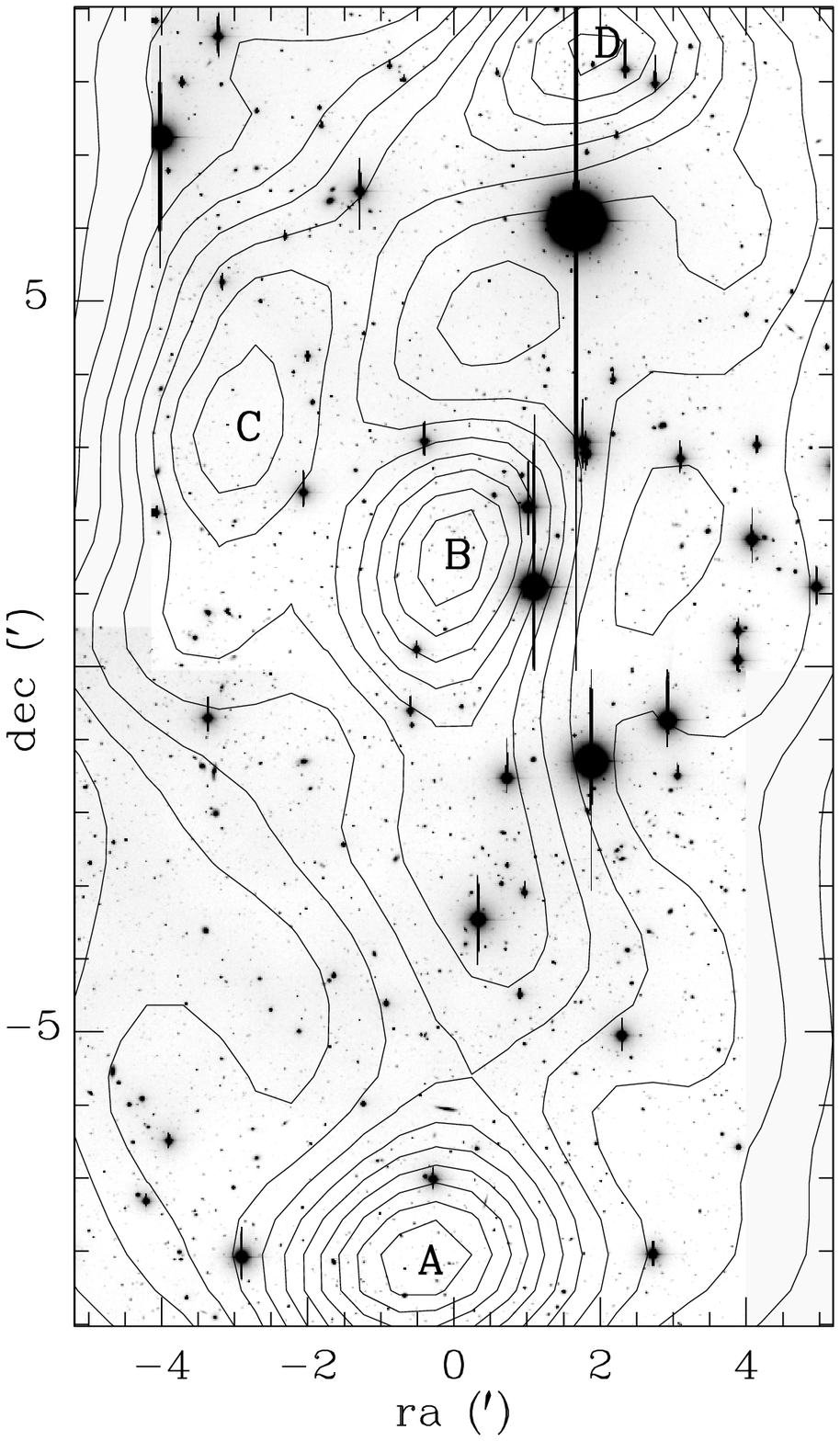}
\caption{Composite $10.4' \times 18.2'$ (4.8 Mpc $\times 8.5$ Mpc) $i$
band image of the supercluster (N is up, E is to the right) Overlaid
are contours of red ($1.2\le R-i \le 1.7$) galaxy density. The
original clusters Cl 1604+4304 (A) and Cl 1604+4321 (D) are at the
bottom and top of this image.  The new clusters Cl 1604+4314 (B) and
Cl 1604+4316 (C) are detected as significant overdensities. Galaxy
density contours range from 0 to 9 galaxies arcmin$^{-2}$, in steps of
0.75 galaxies arcmin$^{-2}$.}
\label{pal}
\end{figure}
\fi

LRIS slitmasks covering parts of this region were observed on UT May
4-7, 2000. A total of six slitmasks were used, with $\sim30$ objects
per mask. Objects with magnitudes down to $i=23.0$ were included.
Data were taken using the 400 l/mm grating blazed at 8500\AA, with a
central wavelength of 7000\AA. This setup provides spectral coverage
from 5500\AA\ to 9000\AA\ and a dispersion of 1.86\AA~pix$^{-1}$; with
the $1''$ slits utilized, the resolution is 7.8\AA\ (550 km s$^{-1}$
at $z=0.9$).  Seeing was typically $0.7''$. Four of the masks were
exposed for $4\times1200$s each, one mask for $3\times1200$s, and one
for $2\times1200$s.  HgNeArKr arc lamps, as well as flat fields, were
observed immediately before and/or after each set of target exposures
to provide wavelength calibration with the instrument and telescope in
the same configuration, minimizing errors due to flexure. The data
were processed using standard IRAF tasks and scripts.

To obtain a larger sample of objects, extending to fainter galaxies,
and covering a wider spatial area, further spectroscopy was obtained
with DEIMOS on UT 25-26 May 2003. Galaxies as faint as $i=24$ were
selected, as well as some objects previously observed with LRIS but
for which no redshift was obtained. We targeted $\sim 85$ objects on
each mask. The instrument was configured with the 1200 l/mm grating,
blazed at 7500\AA. The central wavelength was set to 7700\AA~such that
OII$\lambda 3727$ and Ca H and K $\lambda \lambda 3934,3968$ at the
redshift of the supercluster would be easily visible. Slit widths were
1'', and where possible, slits were aligned with the major axis of the
galaxy. This configuration results in a pixel scale of
0.33\AA~pix$^{-1}$, a resolution of $\sim$1.7\AA\ (120 km s$^{-1}$ in
the cluster frame), and spectral coverage from 6385\AA~to 9015\AA.  We
observed four masks with a total exposure time of 12600s ($7 \times
1800$s) per mask. The large number of exposures permits excellent
cosmic ray subtraction and a significant reduction in background
noise.  Seeing ranged from 0.9'' to 1.5'' on the first night and was
nearly constant at $\sim0.6''$ on the second night. For each mask
setup, we obtained a set of three internal flatfields, as well as an
arc line calibration exposure with the NeArKrXe lamps. Data reduction
was performed using the DEEP2 \citep{dav03} version of the spec2d data
reduction pipeline. Based upon the Sloan Digital Sky Survey (SDSS)
spectral reduction package, this IDL software performs all of the
standard processing steps, producing sky-subtracted,
wavelength-calibrated one- and two-dimensional spectra\footnote{For
details on this pipeline, see the DEEP2 website at {\tt
http://astron.berkeley.edu/$\sim$cooper/deep/spec2d/}}.

Redshifts for the LRIS and DEIMOS observations were all measured using
an IRAF task called {\em redsplot} (T.\ Small, private
communication). This tool allows the user to interactively identify
features in the one-dimensional spectra and displays the locations of
other possible features from a user-provided spectral line list. In
this way, one can centroid various visible features and test different
redshifts to find which produces a consistent set of identified lines.
Median velocity errors for the LRIS and DEIMOS data are 150 and
90~{\kms}, respectively.

The six LRIS masks contained a total of 179 slitlets; of these, 156
were distinct targets. An additional four objects were found
serendipitously for a total of 160 unique objects. Identifications
were made for 147 of these, with 116 (80\%) being extragalactic, and
65 of these (56\%) having redshifts within the supercluster range of
$0.85 < z < 0.96$.  The four DEIMOS slitmasks had 326 unique slitlets.
An additional 57 objects were detected serendipitously, yielding a
total of 383 objects. Of these, we were able to identify 309 objects,
only 22 of which are stars.  Over 40\% (122) of the identified objects
were part of the supercluster complex. The combined LRIS and DEIMOS
data give 378 unique extragalactic redshifts.

Our analysis also includes the original sample of LRIS
spectroscopy measured for Cl 1604+4304 and Cl 1604+4321 (O98). Objects
observed as part of that survey were distributed over an
approximately $2' \times 7'$ region centered on each cluster.
Redshifts were measured for 103 and 135 galaxies in the fields of Cl
1604+4304 and Cl 1604+4321, respectively (P98,P01).

The combined spectroscopic survey yields
a total of 597 unique extragalactic redshifts. In Figure~\ref{zhist}, we show
the redshift histogram for the 289 galaxies in the range $0.8 <
z <1.0$. There are four clear peaks in this distribution which
correspond to the four member clusters of the supercluster. Redshift
peaks at $z = 0.90$ and $z = 0.92$ belong to the original clusters, Cl
1604+4304 and Cl 1604+4321, respectively. The new clusters identified
through the multi-band imaging (see Figure~\ref{pal}) correspond to
the redshift peaks at $z = 0.87$ (Cl 1604+4314) and $z = 0.94$ (Cl
1604+4316).  Table~\ref{clusterdata} lists the name, the reference
letter used throughout the Letter, the central coordinates, and the
median redshift for each cluster. The central coordinates and median
redshift are measured from those galaxies which are counted as members in
the detailed analysis of cluster membership presented in \S 3.

\ifsubmode
\else
\begin{figure}
\plotone{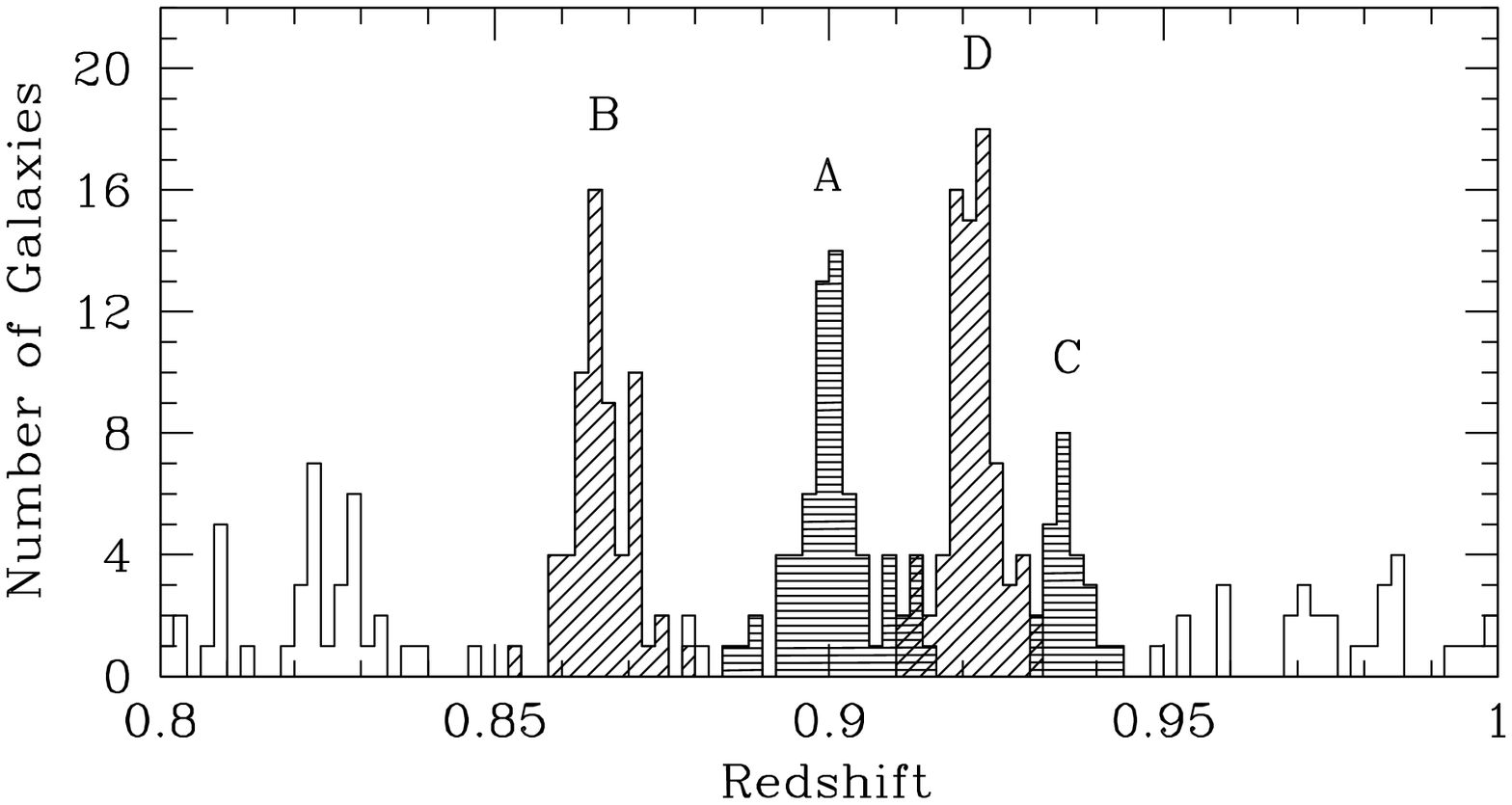}
\caption{The redshift distribution for all galaxies within $0.8<z<1.0$. 
The four clear peaks correspond to the four member clusters in the
supercluster. Each component cluster is shaded and labeled with its
identifying letter (see Table~\ref{clusterdata}).}
\label{zhist} 
\end{figure}
\fi

Figure~\ref{gal_posns} shows the location of all 597 spectroscopic
galaxies, with the 229 confirmed supercluster members coded by cluster
membership. Although the galaxies belonging to each member cluster
appear reasonably concentrated and separated from one another, there
is significant overlap throughout the supercluster region. In
addition, the spatial distribution of galaxies suggests that we have
not yet reached the edge of this structure. We note, however, that the region 
has not been uniformly sampled.

\ifsubmode
\else
\begin{figure}
\epsscale{.70}
\plotone{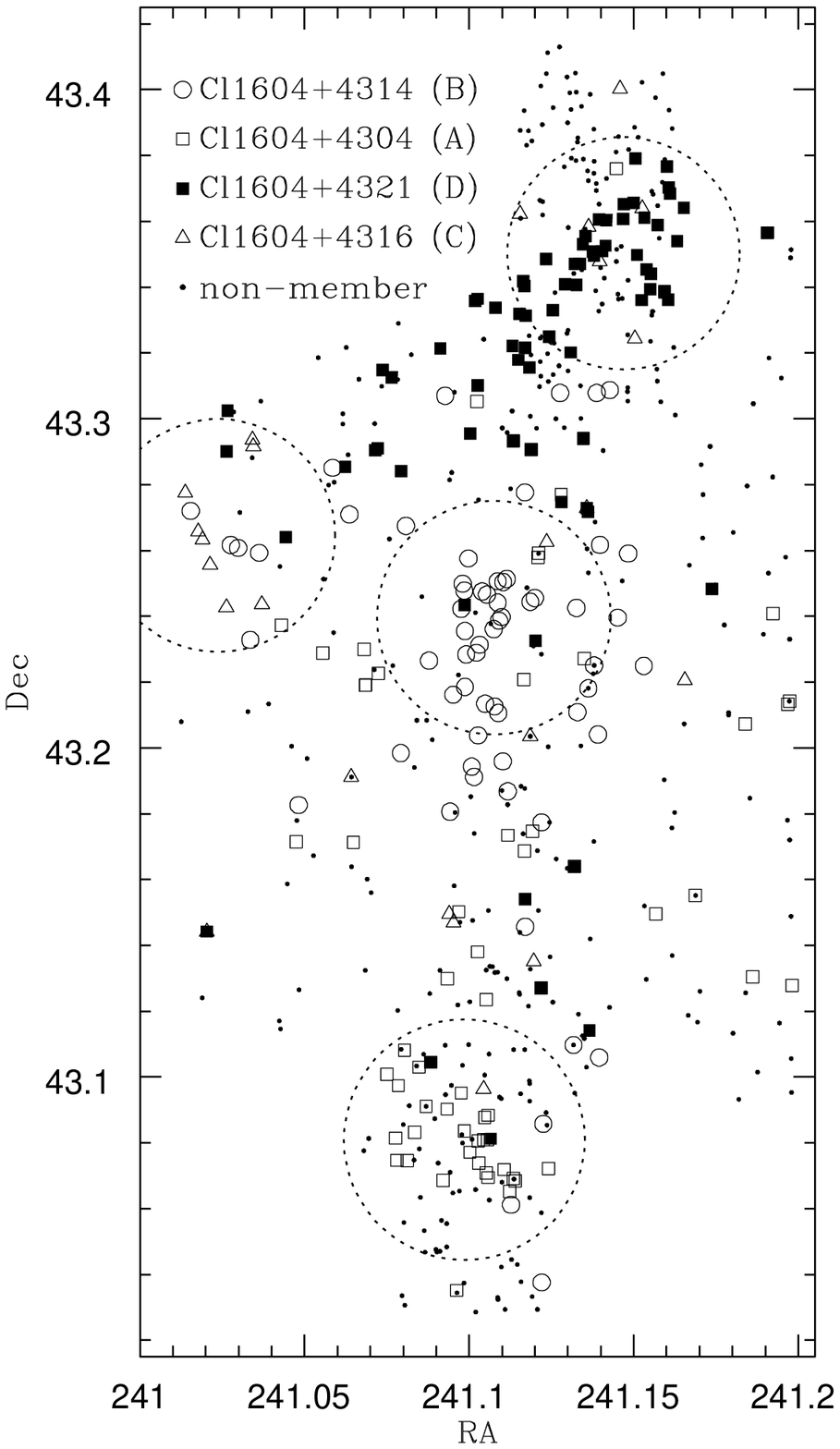}
\caption{The spatial distribution of 597 galaxies with redshifts in
the Cl 1604 supercluster region, using all of the LRIS and DEIMOS
data. Supercluster members are plotted with large symbols, coded by
cluster membership, while non-members are plotted as points. Dashed
circles represent a 1 $h_{70}^{-1}$ Mpc (projected) radius centered on
each cluster.}
\label{gal_posns}
\end{figure}
\fi

\section{Cluster Properties}

\subsection{Velocity Dispersions}

Velocity dispersions are notoriously difficult to calculate for galaxy
clusters, although this has not prevented the publication of results
using as few as ten redshifts per cluster. Selecting cluster members,
a radius in which to include objects, the mathematical technique, and
even the mean cluster redshift each pose their own subtle
issues. These problems are compounded in densely populated areas where
the individual clusters overlap in a larger structure. Nevertheless,
we present initial estimates of the cluster velocity dispersions,
while postponing detailed discussion for a future work, with the goal
of demonstrating that the supercluster constituents are indeed
significant structures, as suggested by P01.

To determine the membership of each supercluster component, we utilize
the ROSTAT package \citep{bee90}, which measures a wide variety of
estimators for the location (median) and scale (line-of-sight
dispersion) of a set of measurements. It avoids assumptions of
gaussianity in the underlying velocity distributions (which is
unlikely to be true for interacting clusters) and includes bootstrap
and jackknife error estimation for the cluster redshifts and
dispersions. Many of these estimators are significantly more resistant
to non-gaussianity in the sample than the traditional Gaussian; large
differences between the different scale estimates can be a sign of
deviations from gaussianity (either in shape or due to outliers) in
the sample.

We first visually divide our redshift histograms into separate (but
overlapping) redshift ranges for each cluster, selecting a
conservative (broad) range for each which clearly includes the entire
structure. The median redshift of each cluster is measured, and
initial, cosmologically corrected peculiar velocities for each galaxy
in the cluster are calculated. These velocities are input into ROSTAT,
which computes the biweight median velocity, and the line-of-sight
dispersions $\sigma_{biwt}$ and $\sigma_{gauss}$, among others.  We
then apply a shift to the velocities of each galaxy such that the
biweight median velocity is zero, while preserving the dispersion.
Objects with $|v|>3\sigma_{biwt}$ are then clipped from the sample,
and the median and dispersions recalculated. This procedure is
iterated until either the various dispersion measures converge, or
there are no more objects to be clipped. For all clusters, the
clipping results in removal of fewer than 10\% of the input
galaxies. The shaded regions in Figure~\ref{zhist} show the final
redshift range for each cluster. Inspection of the spatial
distribution of the clipped galaxies reveals that they are typically
at large clustocentric distances. We note that using only those
galaxies within the central 1.5 Mpc yields comparable dispersions.

\ifsubmode
\else
\begin{deluxetable*}{cccccccll}
\tabletypesize{\footnotesize}
\tablecolumns{9}
\tablewidth{0pc}
\tablecaption{Cluster Properties}
\tablehead{
\colhead{\bf{Cluster}} & \colhead{ID} & \colhead{RA}  & \colhead{Dec} & \colhead{Redshift} & \colhead{$z$ range} & \colhead{$N_{gals}$} & \colhead{$\sigma_{los,biwt}$} & \colhead{$\sigma_{los,gauss}$} }
\startdata
Cl 1604+4304       & A & 16 04 23.7    & +43 04 51.9 & 0.9001 & 0.8852 - 0.9146 & 67 & $962\pm141$ & $989^{+98}_{-76}$ \\
Cl 1604+4314       & B & 16 04 25.8    & +43 14 23.0 & 0.8652 & 0.8539 - 0.8791 & 62 & $719\pm71$ & $726^{+67}_{-51}$ \\
Cl 1604+4316       & C & 16 04 05.7    & +43 15 52.8 & 0.9350 & 0.9302 - 0.9420 & 24 & $489\pm79$ & $479^{+89}_{-57}$ \\
Cl 1604+4321       & D & 16 04 35.3    & +43 21 01.6 & 0.9212 & 0.9111 - 0.9302 & 77 & $640\pm71$ & $649^{+59}_{-46}$ \\[-5pt]
\enddata
\label{clusterdata}
\end{deluxetable*}
\fi

For each cluster, Table~\ref{clusterdata} lists the final redshift
range, median redshift, and number of galaxies, as well as two
estimates of the line-of-sight velocity dispersion (in {\kms}):
$\sigma_{biwt}$, the biweight estimator and its 1-sigma bootstrap
error \citep{bee90}, and $\sigma_{gauss}$, the Gaussian estimator and
its 1-sigma Student's $t$ error.  We note that the bootstrap error
estimates are typically 25\% to 40\% higher than those estimated using
the Student's $t$ distribution, which assumes an underlying Gaussian
population \citep{dan80}. The $\sigma_{los}$ estimates for clusters A
and D are $\sim30\%$ lower than those measured by P01, who used a
factor of $\ge 2$ fewer galaxies. Comparison of our redshift histogram
with their Figure 3 shows that P01 included members of the distinct
structures B and C when measuring the dispersions for A and D,
respectively. Such discrepancies highlight the fact that velocity
dispersions of clusters (and their error estimates) based on small
($\sim20$) redshift samples are subject to strong systematics and
should be regarded with skepticism, especially in the presence of
larger scale structures.

\subsection{Separations}

One goal of this study is to determine whether or not these clusters
are components of a larger system (a supercluster) or a chance
projection on the sky. We estimate the cluster centers (listed in
Table~\ref{clusterdata}) by starting with the position of the peak in
red galaxy density for each cluster (see Figure~\ref{pal}). We then
calculate the median position of all spectroscopically confirmed
members within a radius of $1~h^{-1}_{70}$ Mpc. Under the assumption
that these cluster are no longer expanding with the Hubble flow, we
measure separations, based on these cluster centers, using the angular
diameter distance. The largest transverse separation is
$7.62~h^{-1}_{70}$ Mpc (cluster pair AD), with the closest pair
separated by only $1.85~h^{-1}_{70}$ Mpc (BC) in projection. The
elongation in redshift space implies much greater radial distances,
ranging from 18.1 (CD) to 93.0 (BC) $h^{-1}_{70}$ Mpc.

If we take the largest radial separation as the maximum extent of the
supercluster, its size is not unusual for local superclusters
\citep{bah88} and is consistent with the cluster-cluster correlation
function, which exhibits power on scales as large as $100~h^{-1}$Mpc
\citep{bah83,phg92,bah03}.  The large aspect ratio suggests that we
are observing a filamentary structure nearly face-on, with the
redshift differences too large to be due solely to peculiar
velocities.  Such elongated structures, with typical scales of $\sim
50~h^{-1}_{70}$ Mpc, and as large as $100~h^{-1}_{70}$ Mpc, are seen in large scale structure simulations
\citep{sat98,evr02,kol02}, large galaxy surveys, such as SDSS
\citep{ein03} and APM \citep{ein02}, as well as individual structures
including the Ursa Major \citep{kop01}, Shapley \citep{qui00}, Corona
Borealis \citep{sma98}, Aquarius \citep{car02}, and North Ecliptic Pole \citep{mul01} 
superclusters.

\section{Discussion}

We have obtained the largest spectroscopic sample to date for a
supercluster at $z\sim0.9$. The initial results suggest that the four
component clusters are part of a single, filamentary structure, with
the individual clusters being massive (Abell Richness class 0--2). As
seen in Figure~\ref{gal_posns}, we may not yet have reached the outer
limits of this structure, and future wide-area imaging and additional
spectroscopy will be required to measure its full extent.  Already our
dataset is comparable in size to studies of local superclusters (see
\S3.2). This structure is by far the best studied of the very few
known superclusters at $z>0.3$ \citep{con96,ros99,gav04}.

The spectroscopic data, especially from DEIMOS, are of sufficient
resolution and quality to measure equivalent widths for both emission
and absorption features. A large fraction of the galaxies in the
cluster show evidence for star-formation (most notably strong [OII]
emission). Examining the correlation of such features with galaxy
location will yield insights into the dynamics and history of the
supercluster system and shed light on the role of cluster-scale
physical processes (e.g., shocks and ram-pressure stripping) in galaxy
star formation. Proposed HST imaging and Chandra X-ray mapping of this
structure will allow us to compare galaxy morphological and
spectroscopic properties and correlate optical substructure with the
location and temperature of the hot gas component. These data,
combined with numerical simulations, may be able to predict future
mergers and perhaps even the end-state of this system.  This survey
will provide a view of a young supercluster rivaling the detailed
studies of local structures.

\acknowledgements 
We dedicate this letter to the memory of Bev Oke, an
extraordinary scientist, collaborator, and friend. We thank the
anonymous referee, C. Fassnacht and G. Squires for helpful comments
and contributions. Data presented herein were obtained at the
W.M. Keck Observatory, which is operated as a scientific partnership
among the California Institute of Technology, the University of
California and the National Aeronautics and Space Administration. The
Observatory was made possible by the generous financial support of the
W.M. Keck Foundation. This research was supported in part by STScI
grant HST-GO-08560.05-A.

\ifsubmode
\clearpage
\begin{deluxetable}{cccccccll}
\tabletypesize{\scriptsize}
\tablecolumns{9}
\tablewidth{0pc}
\tablecaption{Cluster Properties}
\tablehead{
\colhead{\bf{Cluster}} & \colhead{ID} & \colhead{RA}  & \colhead{Dec} & \colhead{Redshift} & \colhead{$z$ range} & \colhead{$N_{gals}$} & \colhead{$\sigma_{los,biwt}$} & \colhead{$\sigma_{los,gauss}$} }
\startdata
Cl 1604+4304       & A & 16 04 23.7    & +43 04 51.9 & 0.9001 & 0.8852 - 0.9146 & 67 & $962\pm141$ & $989^{+98}_{-76}$ \\
Cl 1604+4314       & B & 16 04 25.8    & +43 14 23.0 & 0.8652 & 0.8539 - 0.8791 & 62 & $719\pm71$ & $726^{+67}_{-51}$ \\
Cl 1604+4316       & C & 16 04 05.7    & +43 15 52.8 & 0.9350 & 0.9302 - 0.9420 & 24 & $489\pm79$ & $479^{+89}_{-57}$ \\
Cl 1604+4321       & D & 16 04 35.3    & +43 21 01.6 & 0.9212 & 0.9111 - 0.9302 & 77 & $640\pm71$ & $649^{+59}_{-46}$ \\
\enddata
\label{clusterdata}
\end{deluxetable}
\fi

\ifsubmode
\clearpage
\begin{figure}
\epsscale{.90}
\plotone{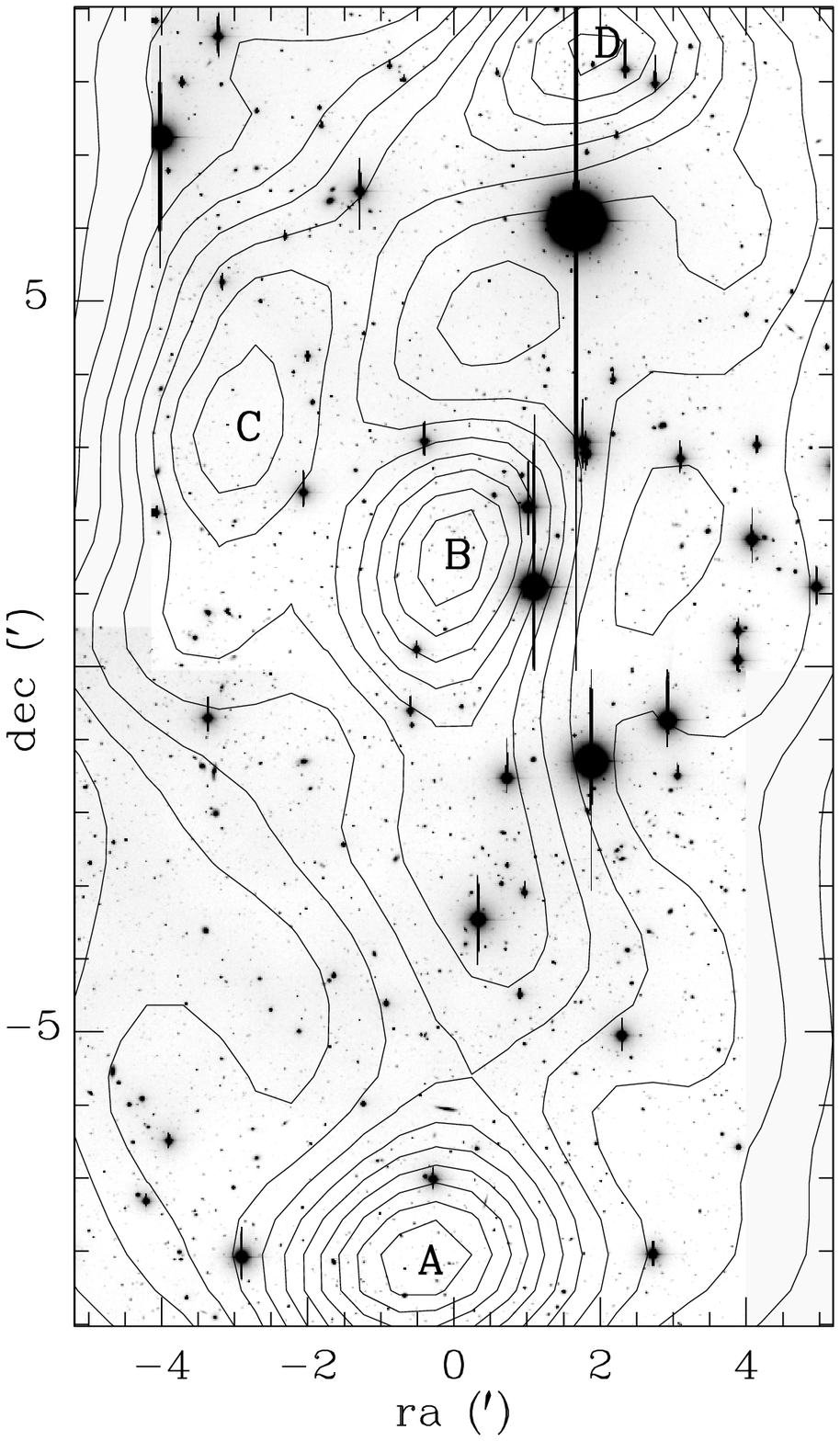}
\caption{Composite $10.4' \times 18.2'$ (4.8 Mpc $\times 8.5$ Mpc) $i$ band image of the supercluster (N is up, E
is to the right) Overlaid are contours of red ($1.2\le R-i \le 1.7$) galaxy density. The
original clusters Cl 1604+4304 (A) and Cl 1604+4321 (D) are at the
bottom and top of this image.  The new clusters Cl 1604+4314 (B) and
Cl 1604+4316 (C) are detected as significant overdensities. Galaxy density contours
range from 0 to 9 galaxies arcmin$^{-2}$, in steps of 0.75 galaxies arcmin$^{-2}$.}
\label{pal}
\end{figure}
\fi

\ifsubmode
\clearpage
\begin{figure}
\plotone{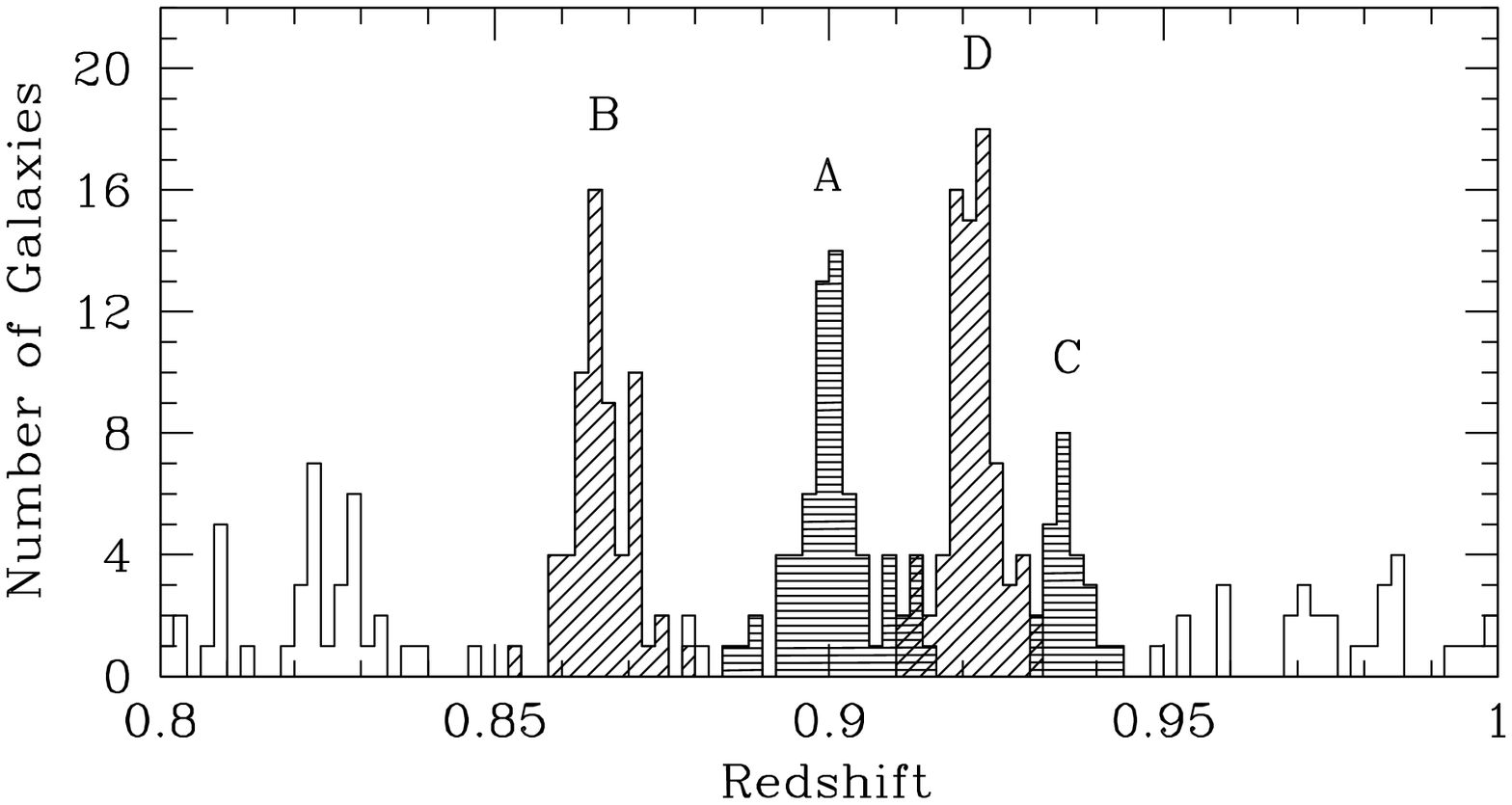}
\caption{The redshift distribution for all galaxies within $0.8<z<1.0$. 
The four clear peaks correspond to the four member clusters in the
supercluster. Each component cluster is shaded and labeled with its
identifying letter (see Table~\ref{clusterdata}).}
\label{zhist} 
\end{figure}
\fi

\ifsubmode
\clearpage
\begin{figure}
\plotone{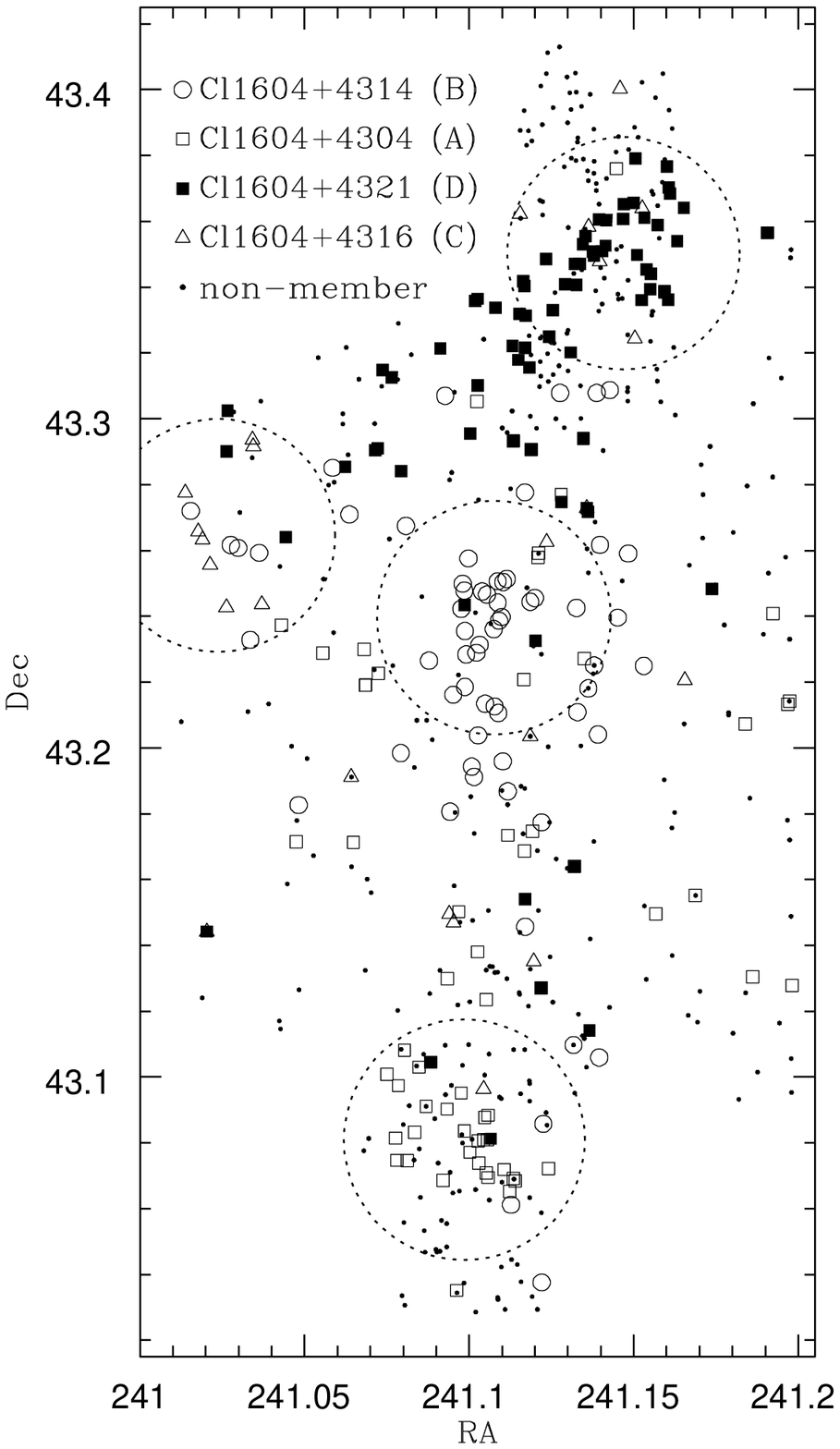}
\caption{The spatial distribution of 597 galaxies with redshifts in
the Cl 1604 supercluster region, using all of the LRIS and DEIMOS
data. Supercluster members are plotted with large symbols, coded by
cluster membership, while non-members are plotted as points. Dashed
circles represent a 1 $h_{70}^{-1}$ Mpc (projected) radius centered on
each cluster.}
\label{gal_posns}
\end{figure}
\fi

\end{document}